\documentclass[twocolumn,english,pra,showpacs,superscriptaddress,longbibliography]{revtex4-1}
\usepackage{amsfonts}
\usepackage{amsmath}
\usepackage{amssymb}
\usepackage[colorlinks, citecolor=blue,linkcolor=red]{hyperref}
\usepackage{color}
\usepackage{float}
\usepackage{hyperref}
\usepackage{textcomp}
\usepackage[rightcaption]{sidecap}
\usepackage{subfigure}
\usepackage[rightcaption]{sidecap}
\usepackage{graphicx}
\begin{document}
\title{Electromagnetically Induced Transparencies with Two Transverse Bose-Einstein Condensates in a Four-Mirror Cavity}
\author{Kashif Ammar Yasir}
\email{kayasir@zjnu.edu.cn}\affiliation{Department of Physics, Zhejiang Normal University, Jinhua 321004, China.}
\author{Zhaoxin Liang}
\affiliation{Department of Physics, Zhejiang Normal University, Jinhua 321004, China.}
\author{Gao Xianlong}
\affiliation{Department of Physics, Zhejiang Normal University, Jinhua 321004, China.}
\author{Wu-Ming Liu}
\email{wliu@iphy.ac.cn}\affiliation{Beijing National Laboratory for Condensed Matter Physics, Institute of Physics, Chinese Academy of Sciences, Beijing 100190, China.}
\affiliation{School of Physical Sciences, University of Chinese Academy of Sciences, Beijing 100190, China.}
\affiliation{Songshan Lake Materials Laboratory, Dongguan, Guangdong 523808, China.}
\setlength{\parskip}{0pt}
\setlength{\belowcaptionskip}{-10pt}
\begin{abstract}
We investigate electromagnetically induced transparencies with two transverse Bose-Einstein condensates in four-mirror optical cavity, driven by a strong pump laser and a weak probe laser. The cavity mode, after getting split from beam splitter, interacts with two independent Bose-Einstein Condensates transversely trapped in the arms of the cavity along $x$-axis and $y$-axis. The interaction of intra-cavity optical mode excites momentum side modes in Bose-Einstein Condensates, which then mimic as two atomic mirrors coupled through cavity field. We show that the probe field photons transition through the atomic mirrors yields to two coupled electromagnetically induced transparency windows, which only exist when both atomic states are coupled with the cavity. Further, the strength of these novel electromagnetically induced transparencies gets increased with an increase in atom-cavity coupling. Furthermore, we investigate the behavior of Fano resonances and dynamics of fast and slow light. We illustrate that the Fano line shapes and dynamics of slow light can be enhanced by strengthening the interaction between atomic states and cavity mode. Our findings not only contribute to the quantum nonlinear optics of complex systems but also provide a platform to test multi-dimensional atomic states in a single system. 
\end{abstract}
\date{\today}
\maketitle

\section{Introduction}
Quantum nonlinear optics -- a quantum interpretation of nonlinear interactions of light at single photon level \cite{Ref1,Ref2,Ref3} -- offers a stunning platform to practically demonstrate information process with quantum computation \cite{Ref4,Ref5,Ref6}. To demonstrate such controlled nonlinear optics, strong and highly nonlinear photon-photon interactions are essential \cite{Ref1}. The electromagnetically induced transparency (EIT) -- a manifestation of quantum interference occurring between two different transitional pathways of photons from a single state \cite{Ref7} -- is proven to be the best example of such nonlinear quantum interactions \cite{Ref8,Ref9}. For this, the coherently driven cavity quantum electrodynamics (cQED) provides an effective foundation to engineer and manipulate the phenomena of quantum nonlinear optics, because of the strong-coupling regime of light and matter \cite{Ref10}. This coherent and strongly coupled environment of the cQED has enabled the researchers to demonstrate the phenomenon of EIT in versatile configurations, like EIT with ionic coulomb crystals \cite{Ref11}, with a single atom \cite{Ref12,Ref13}, and with Rydberg blockades \cite{Ref14,Ref15,Ref16}. The crucial implication of quantum interference occurs when they stop or slow-down \cite{Ref18} the transmitting light in the EIT interval providing a platform for the creation of optical storage devices \cite{Ref4,Ref7}. 

The mechanical characteristics of light in optomechanical system \cite{Ref18,Ref291,Ref29,Ref48,Ref23,Ref49,Ref50,Ref51,Ref52} yield to phonon induced transparencies \cite{Ref19,Ref20}, which can be referred as optomechanically induced transparency (OMIT) \cite{Ref21,Ref22}. The coupling produced by the mechanical effects of light between multiple oscillators, notably mirrors and ultra-cold atomic states, further lead to the concept of multiple EITs \cite{Ref24,Ref25,Ref26,Ref27,Ref28}. These transparencies occur because of quantum interferences in multiple transitional pathways at intermediate states of the system. However, the recent discussions on Fano resonances -- a phenomenon of quantum nonlinear interaction that consequently occurs because of the off-resonant interference \cite{Ref280,Ref2800} -- in a four mirror optomechanical cavity, with two vibrating mirrors, have raised another aspect of engineering hybrid systems \cite{Ref281,Ref282}. Although, the phenomenon of quantum nonlinear optics has been discussed in complex systems \cite{Ref283,Ref284,Ref285,Ref286}, but the demonstration to configure ultra-cold atomic states in such multidimensional hybrid system is essential, especially with respect to quantum nonlinear optics.

In this paper, we investigate EITs with two Bose-Einstein condensates (BECs) transversely coupled to a four mirror cavity, driven by a pump and a probe laser. The high-$Q$ (quality) factor of the cavity generates strong cavity mode which couples both atomic states after getting split from the beam splitter (BS) \cite{Ref281,Ref282,Ref283,Ref284,Ref285,Ref286}. The optomechanical effects of strong cavity mode, after interacting with atomic degrees of freedom, cause the excitation of momentum side-modes in BECs, acting like two atomic oscillators coupled with each other through cavity field \cite{Ref292}. We show that the quantum interference, occurring from both atomic states during probe-cavity transitions, generate two novel EIT windows in cavity transmission and which only exist when both atomic states are coupled with cavity. The coupling strengths of atomic degrees of freedom with system also provide tunability for the Fano line shapes occurring in off-resonant domain. Further, we show the dynamics of fast and slow light induced by transversely coupled BECs and illustrate that the increase in coupling strengths will robustly slow down the transmitting probe light.  

The manuscript is arranged as follows: In Section \ref{sec1}, the detailed system description as well as the mathematical formulation of the considered model are discussed. The Section \ref{sec2} provides the results and discussion about the emergence of EIT in the four mirror cavity with BECs. The Section \ref{sec3} illustrates and explains the behavior of Fano resonance. The section \ref{sec4} contains the results and discussion about dynamics of fast and slow light. Finally, Section \ref{sec5} contains the conclusion of the work.

\section{System Description and Mathematical Modeling}\label{sec1}
\begin{figure}[tp]
	\includegraphics[width=7cm]{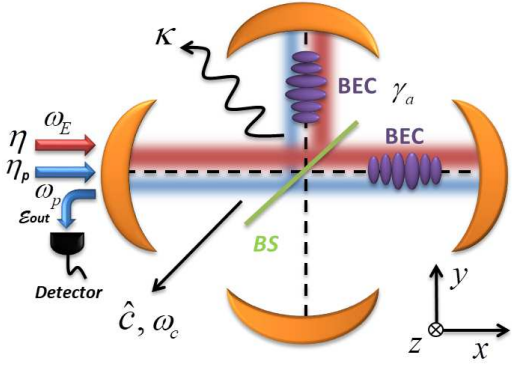}
	\caption{The schematic diagram of a four mirror cavity with two transversely coupled Bose-Einstein condensates, oriented along $x$-axis and $y$-axis. Two lasers, an external pump and a probe, drive the cavity and generate coupling between atomic states after splitting from beam splitter (BS). $\eta$ ($\omega_E$) and $\eta_p$ ($\omega_p$) correspond to the intensities (frequencies) of the external pump and probe lasers, respectively.}
	\label{fig1}
\end{figure}
We consider a four mirror cavity containing transversely located two BECs along $x$-axis and $y$-axis, as illustrated in Fig. \ref{fig1}, unlike the previous investigations where two moving-end mirrors were considered in four mirror optomechanical systems without BECs \cite{Ref281,Ref282}. External pump laser, with intensity $\eta$ and frequency $\omega_E$, and probe laser, with intensity $\eta_p$ and frequency $\omega_p$, longitudinally ($x$-axis) drive the cavity with detuning $\Delta_p=\omega_E-\omega_p$. Due to high quality ($Q$-) factor of the cavity mirrors, the intra-cavity photons perform multiple round trips inside cavity building a strong cavity mode with detuning $\Delta_c=\omega_E-\omega_c$, where $\omega_c$ is the frequency of cavity mode. The cavity mode, after getting equally split from the ($1/2$) BS, interacts with the BECs transversely located in the two arms of the cavity and excites momentum side-modes. These momentum side-modes of BECs can be analogically considered as two atomic mirrors coupled with each other through cavity field. $\kappa$ accommodates the effective cavity mode decay rate, including the photon leakage from the BS to the bottom mirror (oriented along $-x$-axis) of the cavity.  

We consider two one-dimensional BECs having quantized motion along $x$-axis and $y$-axis. By considering large atom-field detuning $\Delta_a$, we eliminate the internal excited state behaviors of BECs, which ultimately leads to the suppression of spontaneous emission. Further by assuming that the BECs are dilute enough, the atom-atom interactions can be ignored. Considering it, we formulated the total Hamiltonian as \cite{Ref22,Ref281}, 
\begin{eqnarray}
\hat{H} &=& \sum_{\sigma=x,y}\int d{\sigma}\hat{\pmb{\psi}}_\sigma^{\dag}(\sigma)\bigg(-\frac{\hbar d^{2}}{2m_{a}d\sigma^{2}}+
\hbar U_{0}\hat{c}^{\dag}\hat{c}\cos^{2}k\sigma\bigg)\hat{\pmb{\psi}}_\sigma(\sigma) \nonumber \\ 
&&+\hbar\Delta_{c}\hat{c}^{\dag}\hat{c}-i\hbar\eta(\hat{c}-\hat{c}^{\dag})-i\hbar \eta_{p}(\hat{c}e^{i\Delta_{p}}-\hat{c}^{\dag}e^{-i\Delta_{p}}).
\label{eq1}
\end{eqnarray}
Here $\hat{\pmb{\psi}}_\sigma(\sigma)=[\hat{\psi}_x,\hat{\psi}_y]^{T}$ ($\hat{\pmb{\psi}}_\sigma^{\dag}(\sigma)=[\hat{\psi}^{\dag}_x,\hat{\psi}^{\dag}_y]^{T}$) represents bosonic field operators for BECs oriented along $x$-axis and $y$-axis, respectively, with equal atomic mass $m_a$. The second term in the Hamiltonian corresponds to the one dimensional optical lattices along $x$-axis and $y$-axis, where $\hat{c}$ ($\hat{c}^\dag$) describes the annihilation (creation) operator for the intra-cavity optical mode. $U_0=g_0^2/\Delta_a$ accommodates the optical potential depth for the atoms defined over the far off-resonant Rabi oscillation $g_{0}$ and atomic-field detuning $\Delta_{a}$ \cite{Ref22}. $k=\omega_E/c$ is the wave number corresponding to the lattice with speed of light $c$. In this study, we assume that the both atomic degrees of freedom are equally interacting coupled to the cavity mode in strong coupling regime $Ng_0^2/\Delta_a>>\kappa$ \cite{Ref31}, where $N$ is the number of atoms in each condensate. The third term corresponds to the intra-cavity optical mode strength while the last two terms accommodate the coupling strengths of intra-cavity field with the external pump laser $\vert\eta\vert=\sqrt{P\times\kappa/\hbar\omega_{E}}$ and probe laser $\vert\eta_p\vert=\sqrt{P_p\times\kappa/\hbar\omega_{p}}$, respectively.

The intra-cavity field interacting with trapped BECs generates photonic recoil causing excitation of symmetrical momentum side $\pm2l\hbar k$, with integer $l$. By assuming weak optical field, i.e. $l=1$, one can consider maximum atomic population saturated in $0^{th}$ and $1^{st}$ mode and can ignore higher-order momentum side-modes. Under this consideration, the bosonic operator for BECs can be formulated as \cite{Ref292},
\begin{equation}
\hat{\pmb{\psi}}_\sigma^{\dag}(\sigma)\approx\frac{1}{\sqrt{L}}[\hat{a}_{\sigma 0}+\sqrt{2}\cos(k\sigma)\hat{a}_{\sigma 1}], (\sigma=x,y).
\end{equation} 
Here $\hat{a}_{x0,y0}$ and $\hat{a}_{x1,y1}$ correspond to the bosonic field operators for atomic distribution in $o^{th}$ and $1^{st}$ momentum side-modes (along $x$ and $y$-axis), respectively. $L=L_x=L_y$ is the effective length of the cavity. By considering modified bosonic operator, the Hamiltonian (\ref{eq1}) reads as \cite{Ref29,Ref30},
\begin{eqnarray}
\hat{H} &=&\sum_{\sigma=x,y} \frac{\hbar U_0}{2}\hat{c}^{\dag}\hat{c}\big(\hat{a}^{\dag}_{\sigma 0}\hat{a}_{\sigma 0}+\hat{a}^{\dag}_{\sigma 1}\hat{a}_{\sigma 1}\big) 
+\frac{\sqrt{2}\hbar U_0}{4}\hat{c}^{\dag}\hat{c}\big(\hat{a}^{\dag}_{\sigma 0}\hat{a}_{\sigma 1}\nonumber \\ 
&&+\hat{a}^{\dag}_{\sigma 1}\hat{a}_{\sigma 0}\big)+\frac{2\hbar^2k^2}{m_a}\hat{a}^{\dag}_{\sigma 1}\hat{a}_{\sigma 1}+\hbar\Delta_{c}\hat{c}^{\dag}\hat{c}
-i\hbar\eta(\hat{c}-\hat{c}^{\dag}) \nonumber \\
&&-i\hbar \eta_{p}(\hat{c}e^{i\Delta_{p}}-\hat{c}^{\dag}e^{-i\Delta_{p}}).\label{eq2}
\end{eqnarray}

As the maximum atomic population is saturated in $o^{th}$ and $1^{st}$ side-modes of BECs, so we can consider $\hat{a}^{\dag}_{\sigma 0}\hat{a}_{\sigma 0}+\hat{a}^{\dag}_{\sigma 1}\hat{a}_{\sigma 1}\approx N, (\sigma=x,y)$, where $N$ is the total number of atoms in each condensate. Further, as the atomic population in $o^{th}$ mode is much higher than the population in $1^{st}$ mode ($N_{0}>>N_1$), therefore, one can approximate that $\hat{a}^{\dag}_{\sigma 0}\hat{a}_{\sigma 0}\approx N$, leading to $\hat{a}^{\dag}_{\sigma 0}$ and $\hat{a}_{\sigma 0}\approx\sqrt{N}$ \cite{Ref292}. By considering these approximation and defining dimensionless position $\hat{q}_\sigma=(1/\sqrt{2})(\hat{a}_{\sigma 1}+\hat{a}^{\dag}_{\sigma 1}), (\sigma=x,y)$ and momentum $\hat{p}_\sigma=(i/\sqrt{2})(\hat{a}_{\sigma 1}-\hat{a}^{\dag}_{\sigma 1}), (\sigma=x,y)$ over the bosonic operator of $1^{st}$ side-modes, with with canonical relation $[\hat{q}_\sigma,\hat{p}_\sigma]=i$, we rewrite the equation \ref{eq2} as,
\begin{eqnarray}
\hat{H} &=&\frac{\hbar U_0N}{2}\hat{c}^{\dag}\hat{c} 
+\sum_{\sigma=x,y}\bigg(\frac{\hbar\omega_\sigma}{2}\big(\hat{p}^2_\sigma+\hat{q}^2_\sigma\big)+\hbar g_\sigma\hat{c}^{\dag}\hat{c}\hat{q}_\sigma\bigg) \nonumber \\ 
&+&\hbar\Delta_{c}\hat{c}^{\dag}\hat{c}-i\hbar\eta(\hat{c}-\hat{c}^{\dag})-i\hbar \eta_{p}(\hat{c}e^{i\Delta_{p}}-\hat{c}^{\dag}e^{-i\Delta_{p}}).\label{eq3}
\end{eqnarray}
Here first term corresponds to the influences of atomic degrees of freedom on the intra-cavity field while the second term describes the motion of atomic side-mode with recoil frequencies $\omega_\sigma=4\omega_{\sigma}=2\hbar k^{2}/m_{a}, (\sigma=x,y)$. The third term defines the coupling of cavity mode with the atomic degrees of freedom, where $g_{\sigma}=\omega_{c}\sqrt{\hbar/m_{bec}\Omega_{\sigma}}/L$ with effective BECs masses $m_{bec}=\hbar\omega_{c}^{2}/(L^{2}NU^2_{0}\Omega_{\sigma})$. The fourth, fifth and sixth terms corresponds to the strength of intra-cavity field and its couplings with external pump and probe fields, as stated previously.

In order to incorporate the effects of dissipation and damping along with the depletion of BECs with standard quantum noise operators, we adopt the quantum Langevin equations approach to govern the collective spatio-temporal dynamics of the subsystems, given as,
\begin{eqnarray}\label{eq4}
\frac{d\hat{c}}{dt}&=&(i\Delta+\sum_{\sigma=x,y}ig_{\sigma}\hat{q}_\sigma-\kappa)\hat{c}+\eta + \eta_{p}e^{-i\Delta_{p}t} \nonumber \\ 
&&+\sqrt{2\kappa} c_{in},\label{2a}\\
\frac{d\hat{p}_\sigma}{dt}&=&-4\omega_{\sigma}\hat{q}_\sigma-g_{\sigma}\hat{c}^{\dag}\hat{c}
-\gamma_{\sigma}\hat{p}_\sigma+\hat{F}_{\sigma},\label{2d} (\sigma=x,y), \\
\frac{d\hat{q}_\sigma}{dt}&=&4\omega_{\sigma}\hat{p}_\sigma-\gamma_{\sigma}\hat{q}_\sigma+\hat{F}_{\sigma}, (\sigma=x,y).\label{2e}
\end{eqnarray} 
Here $\Delta=\Delta _{c}-NU_{0}/2$ corresponds to the effective detuning of the collective system while the $c_{in}$ represents Markovian noise operator associated with cavity input field. The association of external harmonic trap with BECs, which we have ignored in this study so-far, forces atomic momentum side-modes to interact with the optical mode inside cavity, which eventually results in the damping of atomic degrees of freedom \cite{Ref292}. $\gamma_\sigma$ (with $\sigma=x,y$) corresponds to such atomic damping factors. Further, $F_\sigma$ (with $\sigma=x,y$) corresponds to the associated quantum noises with the motion of atomic degrees of freedom, which are assume to be Markovian \cite{Ref32}. Here it should be noted that, unlike macroscopic cavity optomechanics, these damping and quantum noise operators are associated with both position $q_\sigma$ as well as momentum $p_\sigma$ because of the microscopic nature of atomic modes \cite{Ref29,Ref30}. However, in our study, by considering strong cavity mode frequency $\hbar\omega_c>>k_BT$, where $k_B$ is the Boltzmann constant while $T$ is the temperature of the external thermal reservoir, we ignored the effects of associated quantum noises \cite{Ref33}.

In order to include the effects of associated first order quantum fluctuations, we linearize quantum Langevin equations over the steady-states of associated subsystems $\mathcal{\hat{O}}(t)=\mathcal{O}_{s}+\mathcal{\delta \hat{O}}(t)$. Here $\mathcal{\hat{O}}$ is a generic operator for any associated subsystems while $\mathcal{O}_{s}$ corresponds to the steady-state of that operator, which can be easily calculated from quantum Langevin equations by putting time derivative equals to zero. The linearized Langevin equations will read as, 
\begin{eqnarray}
\partial_t\delta\hat{c}(t) =&&-(\kappa+i\Delta)\delta \hat{c}(t)+\sum_{\sigma=x,y}iG_{\sigma}\hat{q}_\sigma+\eta_{p}e^{-i\Delta_{p}t}, \\ 
\partial_t\delta\hat{c}^{\dag}(t) =&&-(\kappa-i\Delta)\delta \hat{c}(t)-\sum_{\sigma=x,y}iG_{\sigma}\hat{q}_\sigma+\eta_{p}e^{i\Delta_{p}t},\\
\partial_t\delta\hat{q}_\sigma(t) =&& 4\omega_{\sigma}\delta\hat{p}_\sigma(t)-\gamma_{\sigma}\delta\hat{q}_\sigma(t), (\sigma=x,y),\\
\partial_t\delta\hat{p}_\sigma(t) =&& -4\omega_{\sigma}\delta\hat{q}_\sigma(t)+G_{\sigma}(\delta \hat{c}(t)+\delta \hat{c}^{\dag}(t)) \\ 
&&-\gamma_{\sigma}\delta\hat{p}_\sigma(t), (\sigma=x,y).
\end{eqnarray}
Here $\partial_t$ represents time derivative and $G_{\sigma}=g_{\sigma}\hat{q}|c_s|$ is the effective coupling between atomic side-modes and the cavity mode, defined by the steady-state of intra-cavity photons.

To measure the behavior of quantum interference in the form of EIT, we have to compute the cavity transmission. In order to do so, we use linearized subsystem quadratures (under condition $\eta>>\eta_p$) in $\delta\mathcal{B}=\sum_{n\rightarrow\{+,-\}}\mathcal{B}_ne^{in\omega t}$, where $\mathcal{B}$ is generic operator for any associated subsystem, and compare the coefficients of probe exponential terms in linearized Langevin equations. After solving these coefficients, the probe field terms appearing with subscription minus $c_{-}$ contains the influences of quantum interference,
\begin{eqnarray}
c_{-}(\Delta_p)&=&\frac{\eta_p(\mathcal{X}_{x}(\Delta_p))}{( \kappa+i(\Delta- \Delta_p))(\mathcal{X}_{x}(\Delta_p)+\mathcal{X}_{y}(\Delta_p))}, \label{eq11}\\
\mathcal{X}_{x}(\Delta_p)&=&1+\frac{G_x^2\omega_x+G_y^2\gamma_x}{\kappa-i(\Delta_p+\Delta)},\nonumber\\
\mathcal{X}_{y}(\Delta_p)&=&\frac{G_x^2G_y^2\omega_y\gamma_y}{(\kappa-i(\Delta_p-\Delta))^2},\nonumber
\end{eqnarray}
where $\mathcal{X}_{x}(\Delta_p)$ and $\mathcal{X}_{y}(\Delta_p)$ can be considered as modified susceptibilities for atomic degrees of freedom along $x$-axis and $y$-axis, respectively, because they contain the corresponding contributions of the condensates.

Further, to compute probe field components in cavity transmission, we use standard input-output relation $c_{out}=\sqrt{2\kappa}c-c_{in}$, which leads to,
\begin{eqnarray}
\mathcal{E}_p(\omega_p)=&&(\eta_p-\sqrt{2\kappa}c_-(\Delta_p))/\eta_p\nonumber \\ 
=&&1-\sqrt{2\kappa}c_-(\Delta_p)/\eta_p.
\end{eqnarray}
Here the amplitude $\mathcal{E}_{out}(\Delta_p)=\frac{\sqrt{2\kappa}c_-(\Delta_p)}{\eta_p}$ defines the absorption (in-phase behavior) and dispersion (out-of-phase behavior) of the  cavity probe transmission with its real and imaginary parts, respectively. Here one can note that the $\mathcal{X}_{y}(\Delta_p)\rightarrow0$ when any of the condensates will be uncoupled to the system (i.e. $G_x\rightarrow0$ or $G_y\rightarrow0$). In this situation, the probe transmission amplitude will read as,  
\begin{eqnarray}
\mathcal{E}_{out}(\Delta_p)=&&\frac{\sqrt{2\kappa}}{\kappa+i(\Delta- \Delta_p)},\label{eq12}
\end{eqnarray}
which is the case of total absorption with no quantum interference (or no EIT) and it can also be seen in the following results as well.

\section{Electromagnetically Induced Transparencies}\label{sec2}
\begin{figure}[tp]
	\includegraphics[width=7cm]{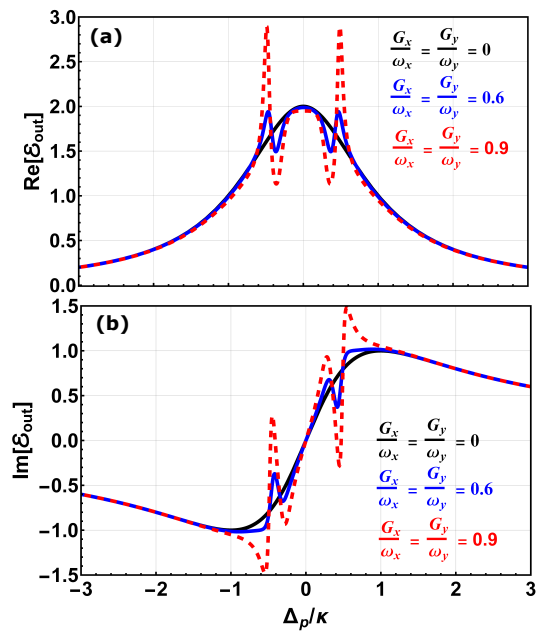}
	\caption{(a) Absorption $Re[\mathcal{E}_{out}]$ and dispersion $Im[\mathcal{E}_{out}]$ of cavity probe transmission as function of normalized pump-probe detuning $\Delta_p/\kappa$, for various atom-cavity couplings $G_x/\omega_x$ and $G_y/\omega_y$. For black, blue and red curves $G_x/\omega_x=G_y/\omega_y=0$, $0.6$, and $0.9$, respectively. The other parameters are considered as, $\Delta=0$, $\gamma_x=\gamma_y=0.1\kappa$, $\omega_x=\omega_y=0.1\Delta_c$, and $\kappa\approx0.1\times2\pi$kHz.}
	\label{fig2}
\end{figure}

The strongly driven cavity mode and the probe laser create double excitation for both transverse condensates (oriented along $x$-axis and $y$-axis). The quantum interference at these double excitation yields to the possibilities for the dark states and leads to the transparencies for probe light \cite{Ref7}. In another analogy, when the intra-cavity optical mode interacts with the condensates, it exerts radiation pressure force. If the radiation pressure force is resonant (or near resonant) with the mechanical frequencies of both condensates, then it generates Stokes and anti-Stokes scattering of probe light from the edges of atomic oscillators. However, under the condition $Ng_0^2/\Delta_a>>\kappa$, the Stokes scattering will be suppressed because the collective system will be in resolved-sideband regime, where the Stokes scattering is off-resonant with cavity mode \cite{Ref7,Ref8,Ref9}. Thus, only anti-Stokes scatterings will survive, which eventually result in the appearance of dark states in probe transmission, as can be seen in absorption ($Re[\mathcal{E}_{out}]$) and dispersion ($Im[\mathcal{E}_{out}]$) of probe transmission illustrated in Figs. \ref{fig2}(a) and \ref{fig2}(b).

In absence of coupling between BECs and cavity ($G_x/\omega_x=G_y/\omega_y=0$), the probe light will be completely absorb by the cavity without facing any quantum interference, as illustrated by the black curves in Figs. \ref{fig2}(a) and \ref{fig2}(b). However, when we couple atomic states with cavity mode, the quantum interference at atomic transitional pathways yields two EIT windows, as can be seen by two dips in blue curves of Figs. \ref{fig2}(a) and \ref{fig2}(b). Further, when we increase the coupling strengths for both condensates, the EIT windows get robustly enhanced, as illustrated by red curves in Figs. \ref{fig2}(a) and \ref{fig2}(b).

The probe light after, getting split from BS, interacts with condensates trapped in two arms of the cavity (along $x$ and $y$-axis). The quantum interaction of probe light at double excitation levels (created by probe light in presence of cavity excitation) of each condensates in each arm of the cavity engineers the feature of EIT for probe light, likewise as it happens in conventional optomechanical systems in the case of OMIT \cite{Ref29,Ref27,Ref28}. After getting reflected from high-$Q$ cavity mirrors in both arms, the probe light with quantum nonlinear feature of EIT merges again at BS, from where it will leave the cavity. Here it should be noted that the quantum nonlinear features of EITs only exist when both atomic condensates are coupled to the cavity mode. If any of the both condensates gets failed to couple with the cavity mode then EIT windows will not appear in the probe transmission. It is because when any of the condensates becomes uncoupled to the cavity field, the probe light from the arm of uncoupled condensate will suppress (dominate) the EIT effects of the probe light coming from other arm with coupled condensate. It eventually results in the absence of EIT windows in probe transmission unlike the case of conventional hybrid optomechanical system, where if any of the oscillators is coupled to the system then there will be an EIT window \cite{Ref29,Ref26,Ref27,Ref28}. 
\begin{figure}[tp]
	\includegraphics[width=7cm]{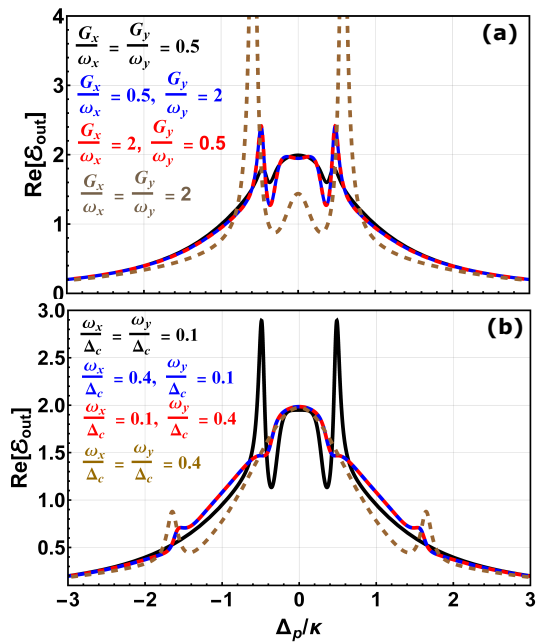}
	\caption{(a) Absorption $Re[\mathcal{E}_{out}]$ versus normalized $\Delta_p/\kappa$, for $G_x/\omega_x=G_y/\omega_y=0.5$ (black), $G_x/\omega_x=0.5,G_y/\omega_y=2$ (blue), $G_x/\omega_x=2,G_y/\omega_y=0.5$ (red), and $G_x/\omega_x=G_y/\omega_y=2$ (brown), at constant $\omega_x=\omega_y=0.1\Delta_c$. (b) Absorption $Re[\mathcal{E}_{out}]$ versus normalized $\Delta_p/\kappa$, for $\omega_x=\omega_y=0.1\Delta_c$ (black), $\omega_x=0.4\Delta_c,\omega_y=0.1\Delta_c$ (blue), $\omega_x=0.1\Delta_c,\omega_y=0.4\Delta_c$ (red), and $\omega_x=\omega_y=0.4\Delta_c$ (black), with constant $G_x/\omega_x=G_y/\omega_y=0.5$. The rest of the parameters are the same as mentioned in Fig. \ref{fig2}.}
	\label{fig3}
\end{figure}

Further, as stated previously, one can mathematically understand this from equation (\ref{eq11}). When any of the condensate is uncoupled $\mathcal{X}_{y}(\Delta_p)\rightarrow0$, which leads to the empty cavity configuration as described in equation (\ref{eq12}). Furthermore, one can also note from the equation (\ref{eq11}) that even in the case stationary $\omega_y=0$ transverse condensate (along $y$-axis) or in the case where it is performing undamped motion $\gamma_y=0$, the EIT effects will also disappear from the probe transmission, because of the suppression of EIT effects coming from longitudinal arm.

In Fig. \ref{fig2}, we discussed the EITs effects when both condensates are equally coupled to the cavity mode. However, in the case where both condensates are not equally coupled to the system, the EITs behavior gets interestingly modified, as illustrated in Fig. \ref{fig3} (a). The decrease in coupling strength of any condensate ($G_x$ or $G_y$) reduces the strength of quantum interference happening in EITs. But, as the both condensates equally contribute to EITs, so if we reduce the strength of coupling for one condensate while keeping the coupling of second condensate constant, or keep the coupling of first one same and reduce the coupling for second one (equal to first one in first case), their influences on EITs will be same. It can be seen in blue ($G_x/\omega_x=0.5,G_y/\omega_y=2$) and red ($G_x/\omega_x=2,G_y/\omega_y=0.5$) curves of Fig. \ref{fig3} (a). 

The increase or decrease in the frequencies ($\omega_x$ and $\omega_y$) of the condensate also equally alter the EITs behavior, as illustrated in Fig. \ref{fig3} (b). By increasing the frequencies of the condensates, the edges of EITs windows move away from the resonant state of pump-probe detuning ($\Delta_p=0$) to the off-resonant domain and mimic like a Fano resonances. It is because higher mechanical frequencies of condensate will prolong the quantum interference by increase the gap between central modes and side-modes of the condensates \cite{Ref29}. Further, in the case of unequal frequencies, the EITs response will be same as it is in the case of unequally coupled condensates, as illustrated by the blue ($\omega_x=0.4\Delta_c,\omega_y=0.1\Delta_c$) and red ($\omega_x=0.1\Delta_c,\omega_y=0.4\Delta_c$) curves of Fig. \ref{fig3} (b).  

\section{Fano Resonances}\label{sec3}
\begin{figure}[tp]
	\includegraphics[width=7cm]{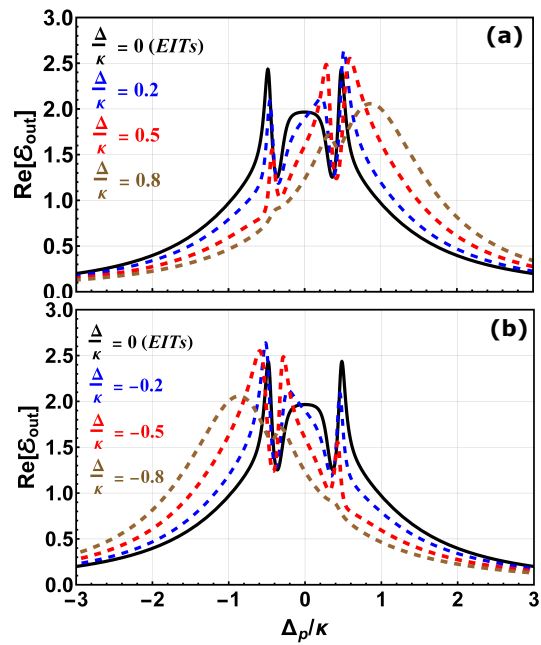}
	\caption{Fano resonance in absorption $Re[\mathcal{E}_{out}]$ versus normalized $\Delta_p/\kappa$ at $G_x/\omega_x=G_y/\omega_y=0.8$. In (a), black, blue, red and brown curves correspond to $\Delta=0\kappa, 0.2\kappa, 0.5\kappa$ and $0.8\kappa$, respectively. In (b), black, blue, red and brown curves correspond to $\Delta=0\kappa, -0.2\kappa, -0.5\kappa$ and $-0.8\kappa$, respectively. Remaining parameters are the same as mentioned in Fig. \ref{fig2}.}
	\label{fig4}
\end{figure}
Fano resonances -- a fascinating implication of quantum nonlinear interactions of light that emerge in off-resonant configuration of EIT -- possess crucial importance in hybrid and complex quantum systems \cite{Ref280,Ref2800}, especially where multiple optical modes are needed to be transmitted through one channel or through one photonic pathway. In optomechanical system the phenomenon of Fano resonances has been extensively studied \cite{Ref29,Ref27,Ref28}, even in the four mirror cavity optomechanical system \cite{Ref281,Ref282}. However, these Fano resonances are worth exploring in our system where two transversely coupled BECs are placed inside a four mirror cavity system. The dynamics of Fano resonances can observed in our system by measuring the probe transmission in off-resonant cavity detuning (pump-caviy) $\Delta$ configuration with respect to the EITs spectrum, as illustrated in Fig. \ref{fig4}. 

When we shift the cavity detuning towards positive ($\Delta>0$) from the resonant (or EITs) regime ($\Delta=0$), the peak of probe transmission starts moving towards the positive pump-probe detuning $\Delta_p>0$, as can be seen in Fig. \ref{fig4} (a). But the EIT dip occurring around $\Delta_p\approx+0.5\kappa$ remain at same position for each increase in cavity detuning $\Delta$ (or for each Fano line shape), yielding in a resonance formation (known as Fano resonance) over the pump-probe detuning $\Delta_p$. It should be noted here, in this configuration, a weak resonance is also appearing around $\Delta_p\approx-0.5\kappa$, but it can be comparatively ignored (or suppressed) with respect to other resonance. Similarly, if we decrease the strength of cavity detuning below zero $\Delta<0$, the probe transmission spectrum (or Fano lines) starts moving towards the left of the pump-probe detuning ($\Delta_p<0$) and the Fano resonance gets sifted to the other EIT dip occurring around $\Delta_p\approx-0.5\kappa$, as illustrated in Fig. \ref{fig4} (b). Here one observe that the width of absorption $Re[\mathcal{E}_{out}]$ increases in off-resonant cavity domain but the Fano resonance will again remain at the same position. The interesting thing that augments the significance of current model is that the both transparencies windows remain at the same position unaffected by the direction system detuning. It is unlike the previous studies \cite{Ref29,Ref28} where the saturation of Fano lines to any EIT window effects the position of second window and presents more appealing argument for the Fano resonance applications. 
\begin{figure}[tp]
	\includegraphics[width=6.5cm]{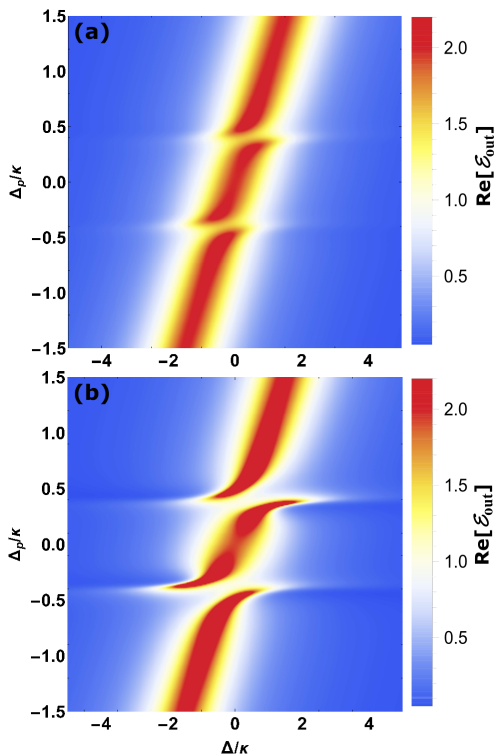}
	\caption{Fano resonance in Absorption $Re[\mathcal{E}_{out}]$ versus normalized pump-probe detuning $\Delta_p/\kappa$ and normalized cavity detuning $\Delta/\kappa$ at $G_x/\omega_x=G_y/\omega_y=0.6$ (a) and $G_x/\omega_x=G_y/\omega_y=1$ (b). Remaining parameters are the same as mentioned in Fig. \ref{fig2}.}
	\label{fig5}
\end{figure}

In order to further explain the behavior of Fano resonances, in Fig. \ref{fig5}, we illustrate the probe absorption $Re[\mathcal{E}_{out}]$ as a function of normalized cavity detuning $\Delta/\kappa$ and normalized pump-probe detuning $\Delta_p/\kappa$. In the absence of BECs-cavity detuning (i.e. $G_x=G_y=0$), the probe transmission will form a diagonal band (bright strip) versus $\Delta/\kappa$ and $\Delta_p/\kappa$ without showing any break or resonance, which is obvious and is not illustrated here. But when we coupled transversely BECs with the cavity arms, two Fano resonances appear in the form of two breaks around $\Delta_p\approx\pm0.5$ in the probe transmission band, as shown in Fig. \ref{fig5} (a). If we further increase the strength of coupling between BECs and cavity, the gap between the beaks will be increased resulting in enhanced Fano resonances, as illustrated in Fig. \ref{fig5} (b). Similarly, like EITs, these dips produced by Fano resonances only exist when both atomic states are coupled to the system and if any of the condensates becomes uncoupled, then there will be no Fano resonance. 

\section{Fast and Slow Light Dynamics}\label{sec4}
\begin{figure}[tp]
	\includegraphics[width=7.5cm]{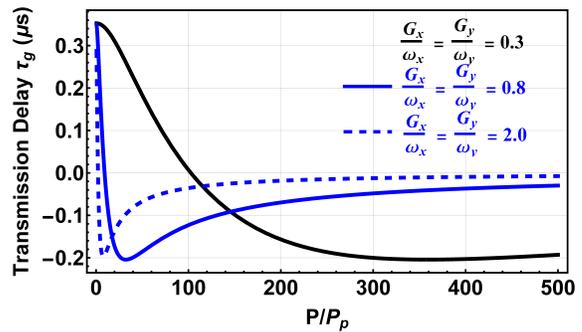}
	\caption{The slow light behavior with the probe group delay $\tau_g$ as a function of external pump laser power $P/P_p$. The black, blue solid, and blue dashed curves represent $G_x/\omega_x=G_y/\omega_y=0.3$, $0.8$, and $2$, respectively. Remaining parameters are the same as mentioned in Fig. \ref{fig2}.}
	\label{fig6}
\end{figure}
The dynamics of fast and slow light possess great significance in order to take quantum nonlinear optical interactions towards practical quantum computation \cite{Ref4,Ref5,Ref6}. In our case, to govern the behavior of slow as well as fast light, the phase $\Phi_p$ of total probe transmission $\mathcal{E}_p$ possesses crucial importance, because with its robust diffusion, one can calculate probe transmission delay (or group delay) $\tau_g$, reading as. 
\begin{eqnarray}
\tau_g=&&\frac{\partial}{\partial\omega_p}\Phi_p(\omega_p)=\frac{\partial}{\partial\omega_p}\bigg(arg\big(\mathcal{E}_p(\omega_p)\big)\bigg)\nonumber\\
=&&\frac{\partial}{\partial\omega_p}\bigg(arg\big(1-\sqrt{2\kappa}c_-(\Delta_p)/\eta_p\big)\bigg).\label{eq13}
\end{eqnarray}
By illustrating the transmission group delay $\tau_g$ versus external pump field power $P/P_p$, we measure the dynamics of fast and slow probe light, as shown in Fig. \ref{fig6}. With weak BECs-cavity couplings $G_x/\omega_x=G_y/\omega_y=0.3$, the group delay decreases for weak external pump laser power $P/P_p$. But the group delay again starts increase with increase in power $P/P_p$ and gets saturated around $\tau_g\approx -0.2 \mu s$ at higher values of $P/P_p$, as can be seen by the black curve of Fig. \ref{fig6}. 
\begin{figure}[tp]
	\includegraphics[width=6.cm]{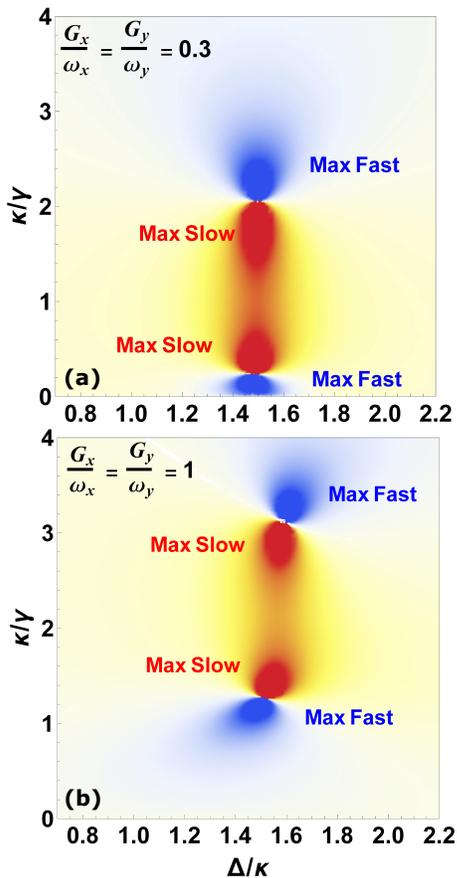}
	\caption{The group delay $\tau_g$ versus normalized cavity detuning $\Delta/\kappa$ and normalized cavity decay $\kappa/\gamma$, where $\gamma=\gamma_x=\gamma_y$, with $G_x/\omega_x=G_y/\omega_y=0.3$ (a) and $G_x/\omega_x=G_y/\omega_y=1$ (b). The shads of blue and red colors corresponds to the distribution of fast and slow light, respectively. Remaining parameters are the same as mentioned in Fig. \ref{fig2}.}
	\label{fig7}
\end{figure}

However, if we increase the coupling strengths of BECs to $G_x/\omega_x=G_y/\omega_y=0.8$, the group delay $\tau_g$ decreases rapidly for initial values of $P/P_p$. But after $P\approx25P_p$, it starts increasing again and this time it saturates at much higher values, around $\tau_g\approx -0.01 \mu s$, for the higher values of $P/P_p$, yielding to slow probe transmission, as illustrated with blue solid curve of Fig. \ref{fig6}. If we further increase the strength of BECs-cavity coupling, the group delay $\tau_g$ will more rapidly decay initially but will get saturated to further higher values with $P/P_p$, as can be seen by the blue dotted curve of Fig. \ref{fig6}, where the BECs couplings have been increase to $G_x/\omega_x=G_y/\omega_y=0.8$. It is happening because, at higher strengths of BECs coupling with cavity, the quantum interference resulting in enhance dark states will increase the widths EITs windows, which eventually yields to slow transmission of probe light. The advantage of current system over the previous studies on fast and slow in atom-cavity systems \cite{Ref28} is the simultaneous effects of both condensates that enhances the group delay of probe transmission.  

To further enhance the understanding of the fast and slow light dynamics, we measured the probe transmission delay versus normalized cavity detuning $\Delta/\kappa$ as well as cavity decay rate $\kappa$, as shown in Fig. \ref{fig7}. It is crucially important to know how much the cavity leakage $\kappa$ can alter the dynamics of fast and slow light, and at which configuration, we can get maximum slow light in our system. At weak BECs-cavity couplings $G_x/\omega_x=G_y/\omega_y=0.3$, the group delay $\tau_g$, between a particular interval of cavity detuning $\Delta/\kappa$, decreases with increase in cavity decay $\kappa$ and reaches at minimum value (maximum fast light) around $\kappa\approx0.2\gamma$ (here $\gamma=\gamma_x=\gamma_y$). However, if we further increase the cavity decay $\kappa$, the $\tau_g$ jumps to its maximum value (maximum slow light) and start decreasing from its maximum value with increase in $\kappa$. But again after $\kappa\approx\gamma$, it starts to increase and reach to the maximum value at $\kappa\approx2.1\gamma$, from where it suddenly jumps to its minimum values, as shown in Fig. \ref{fig7} (a). The higher values of $\tau_g$ or the slow light is almost forming a oval like shape during the cavity decay interval $0.2\gamma<\kappa<2.1\gamma$, where the maxima of slow light are occurring at the edges with the maxima of fast light. However, if we increase the couplings between BECs and cavity, that oval shape will shift towards higher values of cavity decay $\kappa$. Because, at higher couplings, the strength of EIT windows will increase allowing probe transmission to overcome (or handle) higher values of $\kappa$. Although, these novel dynamics of fast and slow light in our system are critically dependent on the various system parameters, but it still magnifies the understanding of the behavior of slow as well as fast light.   

\section{Conclusion}\label{sec5}
We investigate electromagnetically induced transparencies in a four mirror cavity with two Bose-Einstein condensates, trapped along the transverse arms of the cavity, and driven by a strong pump laser and a weak probe laser. The cavity mode, strongly driven by the external pump laser and a weak probe laser, excites atomic states to the double excitation configuration with an intermediate level. We show that the quantum interference occurring at these double excitation levels leads to the two novel transparency windows, which only exist when both Bose-Einstein condensates are coupled to the system. The strength of these electromagnetically induced transparencies can not only be increased with an increase in atom-cavity couplings, but the frequencies of condensates also equally contribute to the transparency windows. Further, by demonstrating the behavior of Fano resonances, we illustrate the occurrence of two Fano resonances corresponding to the two transparency windows and the strength of these windows can be increased with an increase in atom-cavity coupling. Furthermore, we illustrate the dynamics of fast and slow light and conclude that the increase in atom-cavity couplings can significantly slow down the transmission of probe light. These results enhance the understanding of quantum nonlinear optics with hybrid complex systems, especially containing ultra-cold atomic states.

\begin{acknowledgments}
L.Z.X. is supported by  Zhejiang Provincial Natural Science Foundation of China under Grant No. Z21A040009 and the National Natural Science Foundation of China under Grant No. 12074344. G.X.L. acknowledges the support of National Natural Science Foundation of China under Grant No. 11774316. W.M.L. acknowledges the support of National Key R\&D Program of China under grants No. 2016YFA0301500, NSFC under grants Nos. 61835013 and 61775242, Strategic Priority Research Program of the Chinese Academy of Sciences under grants Nos. XDB01020300 and XDB21030300. 
\end{acknowledgments}

\end{document}